# Enhanced magnetic properties in ZnCoAlO caused by exchange-coupling to Co nanoparticles

Qi Feng[1,2], Wala Dizayee[1], Xiaoli Li[1,3], David S Score[1], James R Neal[1], Anthony J Behan[1], Abbas Mokhtari[1,4], Marzook S. Alshammari[1,5], Mohammed S. Al-Qahtani[1,6], Harry J Blythe[1], Roy W Chantrell[7], Steve M Heald[8], Xiao-Hong Xu[3], A Mark Fox[1] and Gillian A Gehring[1]

[1] Department of Physics and Astronomy, The University of Sheffield, S3 7RH, UK
[2] SKLSM, Institute of Semiconductors, Chinese Academy of Sciences, Beijing 100083, P. R. China,
[3] Key Laboratory of Magnetic Molecules and Magnetic Information Materials of Ministry of Education and School of Chemistry and Materials Science, Shanxi Normal University, Linfen 041004, China
[4] Department of Physics, Islamic Azad University Arak Branch, Arak, Iran
[5] The National Centre of Nanotechnology, Materials Science Institute, King Abdulaziz City for Science and Technology, P.O Box 6086, Riyadh 11442, Saudi Arabia
[6] Department of Physics and Astronomy, College of Science, King Saud University, P.O. Box 2455, Riyadh 11451, Saudi Arabia
[7] Department of Physics, The University of York, York, YO10 5DD, UK
[8] Advanced Photon Source, Argonne National Laboratory, Argonne, Illinois, 60429, USA

E-mail: g.gehring@sheffield.ac.uk,  fengqi@iphy.ac.cn

**Abstract**

We report the results of a sequence of magnetisation and magneto-optical studies on laser ablated thin films of ZnCoAlO and ZnCoO that contain a small amount of metallic cobalt. The results are compared to those expected when all the magnetization is due to isolated metallic clusters of cobalt and with an oxide sample that is almost free from metallic inclusions. Using a variety of direct magnetic measurements and also magnetic circular dichroism we find that there is ferromagnetism within both the oxide and the metallic inclusions, and furthermore that these magnetic components are exchange-coupled when aluminium is included. This enhances both the coercive field and the remanence. Hence the presence of a controlled quantity of metallic nanoparticles in ZnAlO can improve the magnetic response of the oxide, thus giving great advantages for applications in spintronics.

# 1. Introduction

The injection of spin-polarized electrons from a ferromagnetic metal into a semiconductor is much less effective than between two semiconductors [1]. This prompts a major research effort to investigate dilute magnetic semiconductors (DMSs), with one of the main goals being to find a semiconductor in which the mobile carriers are spin-polarised at room temperature. Since the ultimate aim is to integrate magnetic storage with logic functions [2,3], we also require that the ferromagnetic semiconductor should have a sizeable magnetization and remanence. These qualities are present in GaMnAs, but at temperatures that are significantly below room temperature [4]. This makes it interesting to explore other magnetic semiconductors with Curie temperatures above 300K, for example those based on oxides.

The study of oxide magnetism has developed very rapidly since the suggestion that doped ZnO should be ferromagnetic at room temperature [5, 6]. Room temperature ferromagnetism has been observed in a range of oxides, ZnO, $SnO_2$, $TiO_2$, $In_2O_3$ when doped with small percentages of transition metals [7]. However, there is much controversy over whether the apparent ferromagnetic signal arises entirely from blocked particles of nanophases, particularly metallic cobalt [8, 9,10,11,]. Moreover, the observation of weak ferromagnetism in undoped oxides [12] raises new questions about the role of defects.

Theoretical models of doped DMS oxides have been built on carrier-mediated interactions between the magnetic transition metal (TM) ion dopants that are added to induce the ferromagnetism. The carriers can be either itinerant or bound to defects in magnetic polarons [13, 14] and ion-polaron coupling has been demonstrated explicitly in colloidal Mn-doped ZnO particles [15]. However, this picture has been challenged by a number of papers based on XMCD [16], Mössbauer [17] and Andreev [18] studies that have shown that the moments on the TM ions do not participate in the ferromagnetism although the sample is ferromagnetic. Furthermore, other workers failed to find any ferromagnetism in epitaxial doped films [19,20] and it has also been suggested that the magnetism resides on the grain boundaries [17, 21].

Aside from the controversy about the origin of the ferromagnetism, potential applications motivate research into ways to improve the magnetic properties, most notably the coercive field and remanence which, at the present time, are invariably too small to be usable [7,17]. In this paper, we demonstrate how to obtain a ZnO-based ferromagnet that possesses both a large density of spin-polarized mobile carriers and a usable remanence. We have achieved this by employing metallic Co nanoparticles to enhance the magnetic hardness of ZnCoAlO through the phenomenon of exchange-coupling. This contrasts with work by other groups in which the aim was to grow films *without* Co nanoparticles so as to be sure to observe true oxide magnetism [22, 23, 24]. We show that the presence of Co nanoparticles is in fact extremely beneficial, provided that they are exchange-coupled to robust magnetism within the oxide matrix. We compare a film containing very little metallic cobalt with two films that both contained metallic cobalt one with and without codoping with Al. Earlier work had shown that adding Al enhanced both the magnetization and the magneto-optic response [13, 25] via ferromagnetic coupling of Co nanoparticles through the ZnAlO matrix [26].

The paper is organized as follows. After describing the experimental methods in Section 2, we first present the results obtained by magnetization measurements in Section 3. Then, in Section 4, we present magneto-optical measurements. In Section 5 we discuss the implications of our results and present a justification of our exchange coupling model. Finally, in Section 6 we give our conclusions.

**2. Experimental methods**

The samples were grown by pulsed laser deposition (PLD) on c-cut sapphire substrates purchased from *Crystal GmbH Berlin*. Before deposition, the substrates were cleaned ultrasonically in methyl alcohol and did not show any substantial ferromagnetism [27]. Powders of ZnO and $Co_2O_3$ and, where desired $Al_2O_3$, were first mixed in the proportion to give the desired Co:Zn ratio and were then ground together using a pestle and mortar for 30 minutes; this method was found to give a better mix than mechanical grinders. In each case the process of first grinding and then sintering the mixture in air for 12 hours was repeated for annealing temperatures of 400 °C, 600 °C, and 800 °C, before final pressing into a target mould and sintering at 1000 °C for more than 12 hours; however the final sintering of the target for sample B was performed at a higher temperature (~1150 $^0$C) so that the target colour changed became a dark green unlike the others that were a light grey green indicating that a change in chemical bonding had occurred in this anneal.. Samples A and B, of thicknesses 131nm and 136nm respectively, were made from a target that contained 10% cobalt and were grown at base pressure, $3\times10^{-2}$ mTorr, with substrate temperature of 400 $^0$C.

Sample C was co-doped with 0.6% Al and was grown in an atmosphere of 10 mTorr oxygen and had a thickness of 140 nm. A Lambda Physik LEXTRA 200 XeCl excimer laser with an operating wavelength of 308 nm and a 10 Hz repetition rate was used for the ablation of the target. The concentration of Co in films A and B was measured by energy dispersive x-ray analysis (EDX) and found to be 20.1% and 14.5% of Co respectively and that for film C was ~25% as measured by particle-induced X-ray emission (PIXE) measurements [23].

The relative amounts of metallic cobalt and $Co^{2+}$ ions were determined by x-ray measurements made on beamline 20-ID-B (samples A and B) and beamline 20-BM-B (sample C) at the Advanced Photon Source as shown in figure 1. Pre-edge measurements on Samples A and B, comparing them to both pure metallic Co and also a sample of ZnCoO that did not contain any metallic cobalt [22], showed that in Sample A 7% of the Co was in a metallic environment in contrast to that on sample B where only a very small fraction of the Co ions were in metallic cobalt. The data for sample C also showed that the fraction of Co ions in metallic Co was small and this was confirmed by EXAFS where the fraction was found to be ≤ 2% [23]. The details of the three samples are given in Table I. In each case the magnetisation that should be expected from the metallic cobalt clusters at 5K is calculated assuming that all the Co ions in the cluster carry the usual magnetic moment for bulk metal of 1.7 $\mu_B$ per ion and the results are also included in Table I.

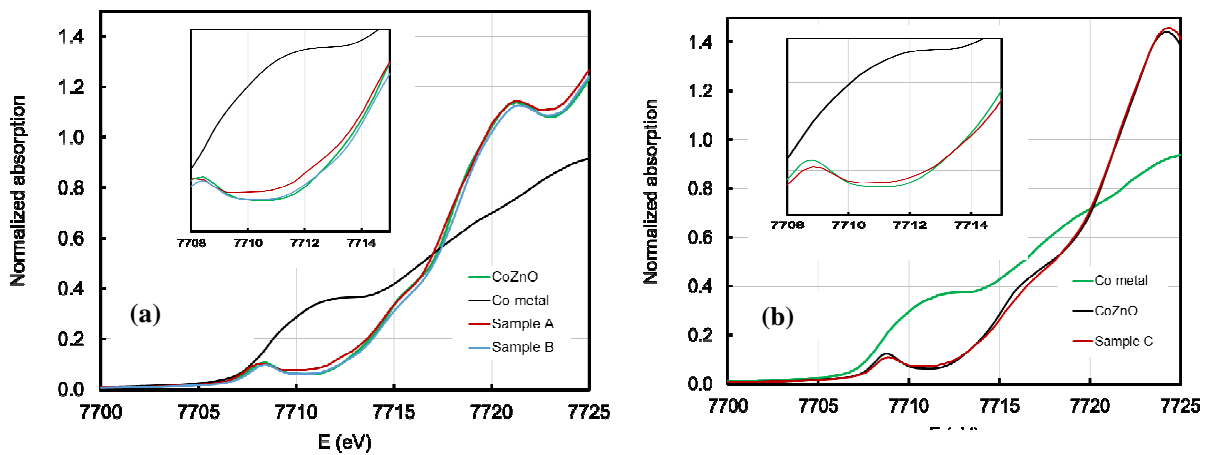

**Figure 1.** The pre-edge region of the Co K-edge for a) samples A and B compared to spectra from a Co doped ZnO film with no metal content and also a sample of metallic Co. The region near 7712 eV is most sensitive to metal. The measurements were made with the x-ray polarization along the c-axis of the ZnO film. b) The equivalent measurements for sample C that were made with the x-ray polarization in the a-b plane of the ZnO film. The insets show an enlarged plot of the most important regions.

A SQUID magnetometer was used to measure the hysteresis loops and the field cooled (FC) and zero field cooled (ZFC) magnetisation. The results shown have all been corrected for the presence of the diamagnetic substrate by measuring a blank substrate and subtracting off its contribution from the total. Magneto-optical data were obtained in Faraday geometry using a photo-elastic modulator over an energy region of 1.5eV-3.4eV. Since this energy range is below the band gap of ZnO at ~3.4eV, magneto-optical responses probes the magnetic polarisation of any gap states. The imaginary part of the off-diagonal dielectric constant was deduced from the measured magnetic circular dichroism (MCD) according to:

$$MCD = \frac{2\omega L}{cn} \operatorname{Im} \varepsilon_{xy}^{eff} \qquad (1)$$

where $\omega$ is the angular frequency, $L$ is the thickness of the film and $n$ its refractive index. Further experimental details are given elsewhere [13, 26, 28].

Three types of MCD measurements were taken: a spectrum that was obtained by subtracting the measured ellipticity in a large negative field from that taken in a large positive field; a remanence spectrum that was taken at zero field after first saturating the sample at large field and a hysteresis loop at a definite energy. In the cases of the spectra taken in field and the MCD hysteresis loop equivalent measurements taken from a blank substrate were subtracted from the data that is presented here.

### 3. Magnetization measurements

This section describes detailed magnetisation measurements made on the three samples: sample A where the magnetization is dominated by the contribution from metallic cobalt, sample B where very little metallic cobalt was detected by x-ray spectroscopy and is a good example of a sample whose magnetisation is due to ZnO doped with $Co^{2+}$, and sample C that shows some effects that are characteristic of both samples A and B. The focus of this paper is to discuss sample C but we discuss samples A and B first as they exemplify the behaviour expected when the samples is and is not dominated by nano-inclusions of metallic cobalt.

**Magnetization of sample A**

Sample A has behaviour that is characteristic of a magnetic sample dominated by cobalt nanoparticles [29, 30]. At low temperatures, where these particles are blocked, they give rise to the large value of the coercive field, 800 Oe, as shown in figure 2(a). This is confirmed by the FC/ZFC cooled magnetisation plots shown in figure 2(d), where a blocking temperature is clearly visible at 30K, which is indicative of blocking of ~3.4nm nanoparticles of cobalt metal below that temperature. Above 30K, there is no discernible difference between the FC and ZFC magnetisations, and the magnetisation may be fitted by a Curie-Weiss plot, $M = \frac{CH}{T+\theta}$, where C=4.2×10$^{-6}$ emu deg$^{-1}$ and $\theta$=66 K, indicating antiferromagnetic coupling between the nanoparticles. The ratio of the remanence to the saturation magnetisation at 5K is close to 0.5 which is the value expected for an isotropic array of nanoparticles with uniaxial anisotropy. However the saturation magnetisation observed at 5K is larger than that expected to result from metallic cobalt alone, as calculated using the measured values of the total cobalt concentration and the fraction of Co ions in a metallic environment (see Table I). Moreover, the coercive field did not drop to zero above the blocking temperature as expected if the magnetisation was dominated by superparamagnetic nanoparticles [9,29,30]; these results together suggest that even for this sample there is also a magnetic contribution from the ZnO doped with 85% of the Co$^{2+}$ ions.

We can obtain further insight by analysing the low field susceptibility for sample A. This quantity can extracted either from the ZFC magnetisation via $\chi_{zfc} = M_{zfc}/H$ or from the hysteresis curve directly using $\chi_{loop}(T) = (\partial M/\partial H)_{H=-H_c}$. Both plots are shown in figure 3(a) and are in good qualitative agreement. These experimental curves are compared to a model that assumes that the observed hysteretic magnetism is entirely due to blocked nanoparticles, and hence that the low field susceptibility is related to the temperature dependence of the remanence [31-34]. In this model some of the nanoparticles, $N_{para}(T)$ will be unblocked below $T_B$ and hence behave superparamagnetically; the remaining particles will be blocked. The low-field susceptibility is dominated by the part of the magnetisation, $N_{para}(T)M_{cluster}$, that is due to superparamagnetic particles with average magnetization, $M_{cluster}$, while the remanence is related to the fraction of the nanoparticles that are blocked. Assuming that the superparamagnetic particles obey a Curie law and that the blocked particles have an average cluster magnetization $M_{cluster}$ that follows Stoner-Wohlfarth theory which relates the remanence to half the saturation magnetization [29,31, 32], we then find:

$$\chi_{cal}(T) = N_{para}(T)\frac{M_{cluster}^2 \mu_0}{3k_B T} = \mu_0 M_{cluster}\frac{2(M_r(5K) - M_r(T))}{3k_B T}. \qquad (2)$$

The qualitative agreement between the experimental data for the zero-field susceptibility and the model is apparent in figure 3(a). In particular the experimental peak at the blocking temperature, 30K, is well reproduced by this model.

The temperature dependence of the coercive field and the remanence are shown in Fig 4(a) and the data agree well with theory from a model of magnetic non-interacting nano-particles given in the Appendix, thus confirming that the magnetic properties of this film are well described by assuming that the magnetisation is due to non-interacting magnetic nanoparticles. Thus all the data on sample A are consistent with the magnetisation being predominantly due to the metallic cobalt nanoparticles that act independently.

**Magnetization of sample B**

Sample B has an overall Co concentration of 14.5% and its saturation magnetization is particularly high although there was hardly any presence of metallic cobalt detected from the near edge absorption measurement of the EXAFS. It is therefore an example of a sample with negligible metallic cobalt. Consequently, its magnetic properties are very different to those of sample A. The shape of the hysteresis loops at 300K are almost unchanged from those taken at 5K (see figure 2(b)) with only a small reduction in the saturation magnetisation at room temperature as given in Table I; the remanence is also low.

**Table 1 Summary of structural and magnetic measurements**

| Sample | Thickness nm | % Co in film | % of Co in film present as metallic cobalt | Saturation magnetisation $M_s$ emu/cm$^{-3}$ at 5K due to metallic Co | Saturation magnetisation $M_s$ emu/cm$^{-3}$ 5K    300K | Coercive field $H_c$ Oe 5K    300K | Ratio remanence to saturation ratio at 5K |
|---|---|---|---|---|---|---|---|
| A | 131 | 20 | ~7 | ~9 | 13±1    9±1 | 800±50    165±50 | 0.5±0.1 |
| B | 136 | 14.5 | Very little | ~1 | 49±5    40±5 | 100±50    50±50 | 0.2±0.1 |
| C | 140 | ~25 | ~2 | ~3 | 15±1    10±1 | 600±50    100±50 | 0.45±0.1 |

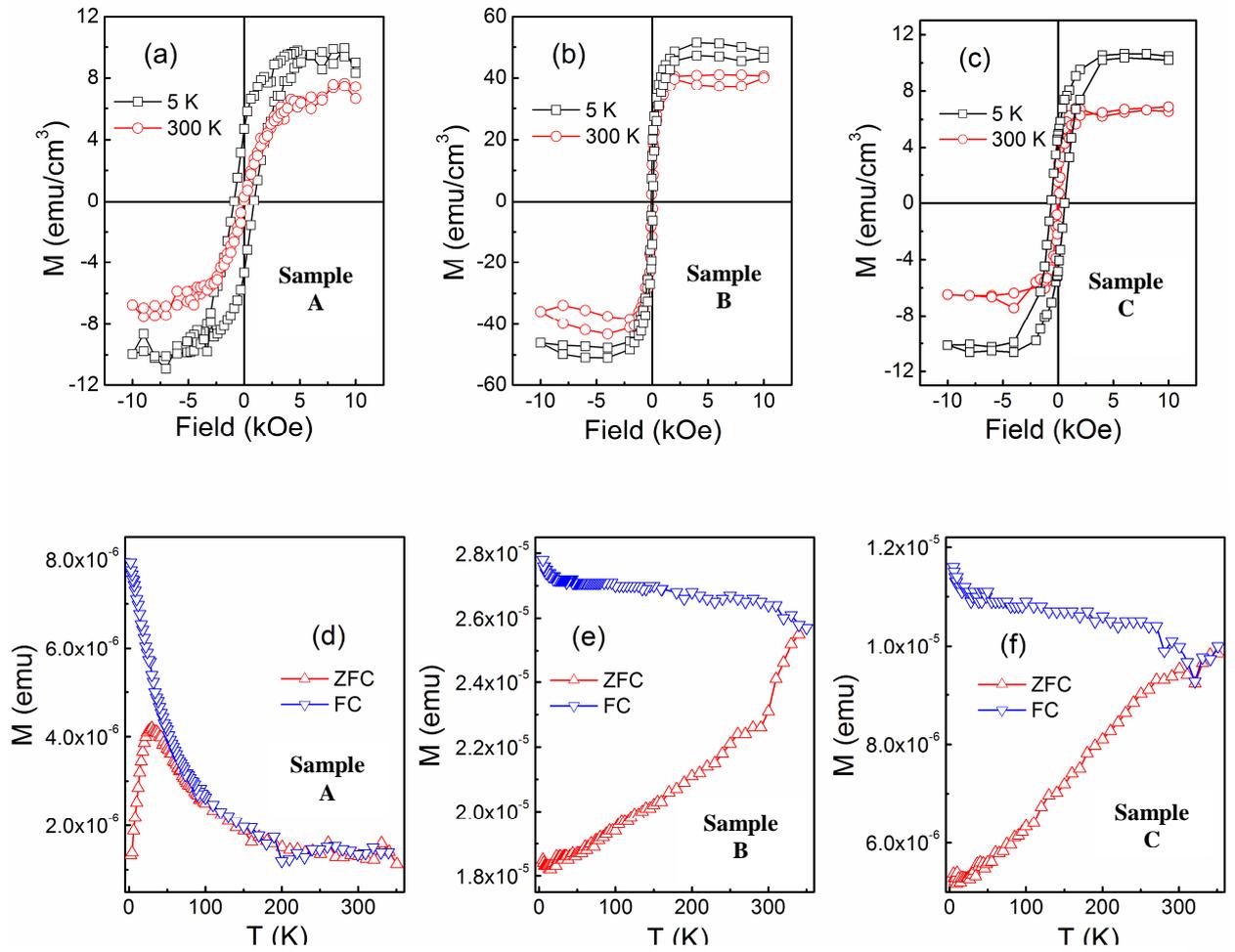

**Figure 2**. (a) (b) and (c) Show the hysteresis loops at 300K and at low temperature for samples A, B and C respectively where the contribution of the diamagnetic substrate has been subtracted. The zero-field-cooled and field-cooled magnetizations are shown in figures (d), (e) and (f). Samples A and B were measured in 100 Oe and sample C in 200 Oe.

The coercive field is very small at all temperatures; however, hysteretic behaviour is seen in the FC/ZFC plot shown in figure 2(e) in the applied field of 100 Oe over the whole temperature range. There is no region in which the ZFC magnetisation plot has the inverse temperature dependence characteristic of a Curie Weiss plot; hence we conclude that in this case the FC/ZFC magnetisation has been obtained entirely within the ordered, ferromagnetic, state. The weak kink in the ZFC plot near the Neél temperature for bulk CoO, namely 291K, might indicate that the sample contains a small amount of antiferromagnetic CoO (too small to be clearly detected by EXAFS). This does not affect the main conclusion that the sample is in an ordered ferromagnetic state over the entire temperature range studied.

**Magnetisation of sample C**

Sample C contains 25% cobalt, with 2% of the Co in the metallic state. This sample also contains 0.6% Al, which has been shown to give a ferromagnetic exchange between cobalt nanoparticles in ZnAlCoO, unlike the antiferromagnetic exchange found between Co nanoparticles in ZnCoO [26]. The magnetic properties of this film are shown in figures 2(c) and 2(f) and have some features in common with both A and B. The large value of the coercive field and the remanence at low temperatures followed by a greatly reduced values at 300K (see table I) are similar to those of film A. However, there is no peak in the FC/ZFC plot corresponding to a blocking temperature and no region below 350K where a Curie Weiss behaviour was detected. The low temperature value of the saturation magnetisation ~15 emu/cm$^3$ was also considerably larger than that expected from metallic cobalt alone, ~3 emu/cm$^3$.

Hysteresis loops were measured at a range of temperatures using SQUID magnetometry. As the temperature is increased the coercive field drops smoothly and can be fitted by a model of a distribution of non-interacting nanoparticles with uniaxial anisotropy as is shown in figure 4; the details of the model, as discussed for sample A, are in the Appendix. However the same model fails to fit the measured temperature dependence of the remanence. The experimental data is unlike anything that is normally given by microscopic modelling as it is normal for the remanence to fall more slowly than the coercive field at low temperatures whereas in this data the initial rates of change are very similar [31].

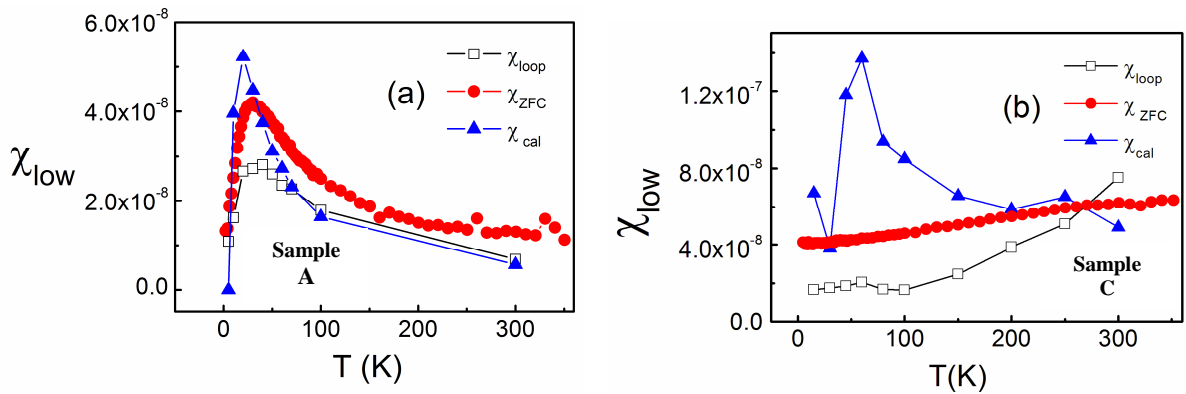

**Figure 3.** Low-field susceptibilities (a) for sample A and (b) for sample C. Black open squares, as measured directly from the hysteresis loop; red circles, from the ZFC plot; blue triangles, as calculated from the measured values of the remanence using equation (2).

Plots of the various expressions for the low field susceptibility for sample C are shown in figure 3(b). The calculated value taken from data shown in figure 4(b) shows a strong peak at ~60K; however no peak is found near this temperature in susceptibility taken from the ZFC magnetisation data nor in the susceptibility found from the hysteresis loops.

## 4. Magneto-optic spectroscopy

The imaginary part of the off-diagonal dielectric constant, $\text{Im}\,\varepsilon_{xy}(\omega)$, may be deduced from MCD spectra using equation (1). This is a very interesting quantity because it gives information about the polarisation dependence of the absorption at each individual energy. The MCD signal is found by measuring the total ellipticity induced by the sample after polarising in a direction along the light path and then subtracting the measurement taken after the sample is polarised in the opposite direction. This measurement is particularly important in samples that contain metallic cobalt and spin polarised ZnO because their contributions dominate at different energies.

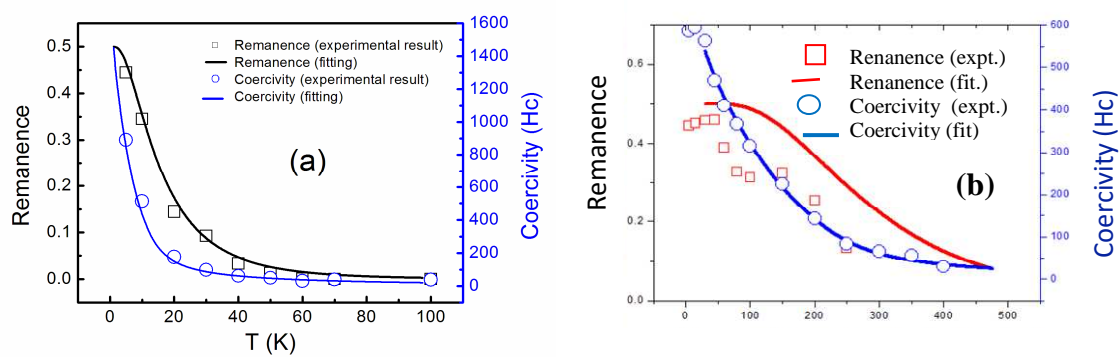

**Figure 4.** The temperature variation of the coercive field, $H_C$, (triangles) and the reduced remanence, $M_r/M_s$, (squares) for samples A in (a) and C in (b). The symbols are data derived from SQUID magnetometry. The curves are obtained from the theoretical model of independent Co particles this is presented in appendix A with parameters chosen to fit $H_c$.

The cobalt nanoparticles are known to give a disproportionably large contribution to the MCD relative to their magnetisation [35] and are discussed first. The Maxwell–Garnett (M–G) theory has been used extensively to analyse Co nanoparticles in non-magnetic hosts [36, 37, 38]. We have extended the theory to the case in which both the nanoparticle inclusion, Co, and the host ZnCoAlO are magnetic therefore contribute to the MCD through $\varepsilon_{xy}^{Co}, \varepsilon_{xy}^{ZO}$. In this case, the effective dielectric function for light propagating along the direction of magnetization, $z$, is given by,

$$\varepsilon_{xy}^{eff} = \varepsilon_{xy}^{ZnO} + \frac{f(\varepsilon_{xy}^{Co} - \varepsilon_{xy}^{ZnO})}{\left[1 + (1-f)\frac{L_{xx}}{\varepsilon_{xx}^{ZnO}}(\varepsilon_{xx}^{Co} - \varepsilon_{xx}^{ZnO})\right]^2}, \qquad (3)$$

where $f$ is the fraction of the sample occupied by the Co inclusions, and $L_{xx}$ is the demagnetising factor of the metallic inclusions. For energies below the band gap we can replace the diagonal part of the dielectric tensor of ZnCoAlO, $\varepsilon_{xx}^{ZnO}$, by the (energy dependent) refractive index squared, $n^2$ [39].

Inclusions of Co that are spherical on average would produce $L_{xx} = 1/3$. However, there is a report [11] of a film grown with 30% Co that contained some spherical inclusions, but also a small fraction of needle-like inclusions growing along the $z-$ axis, giving rise to an average value of $L_{xx} = 0.41$. Hence, in what follows, we treat $L_{xx}$ as a fitting parameter. The real and imaginary parts of the dielectric constants of bulk Co metal, $\varepsilon_{xx}^{Co}$ and $\varepsilon_{xy}^{Co}$, are known, but they are not appropriate for Co nanoparticles. We therefore follow the practice of fitting the bulk dielectric constants to a Drude model [40] and then treating the relaxation time, $\tau$, as an adjustable parameter [36, 37, 38, 41]. The value of $\text{Im}\,\varepsilon_{xy}(\omega)$ is negative for low energy and then crosses zero at an energy that depends on $L_{xx}(1-f)$, but is between 2.5 eV and 3.5 eV [41]. We note that in the fitting data to equation (3) there are three parameters, $f$, $L_{xx}$ and $\tau$. The overall strength of the signal is dominated by $f$, the crossing point by $L_{xx}$ and the width of the curve by $\tau$ thus aiding the determination of each parameter separately [41].

The spectrum of a sample dominated by magnetism in the oxide due to a polarised donor band is very different from the M–G spectrum and is characterised by a strong dip near the band gap at ~3.4eV [13,42,43]. The relative size of the band edge signal and the response due to metallic Co nanoparticles enables us to quantify the relative strengths of the oxide and metallic cobalt in each sample.

The MCD spectrum of sample A, taken in field, is fitted very well by equation (3) as shown in figure 5(a) where the value of $\text{Im}\,\varepsilon_{xy}^{eff}$ obtained from equation (3) is compared with that obtained from the MCD using equation (1). The fitting parameters are given in Table II.

The fraction of the volume of the film that is occupied by metallic Co was found to be $f_{MCD} = (3\pm1)\times10^{-3}$ which may be compared with what is expected from the measurements of the Co concentration and the fraction of Co ions that have a metallic environment, as obtained from x-ray data and given in Table I, which gives $f_{struct}= (7\pm1) \times10^{-3}$. The larger value obtained from the x-ray data implies that not all the Co atoms that have metallic valence are contributing the full magneto-optic response of bulk cobalt. There is a slight deviation close to the band edge that may be coming from a polarised defect in the oxide and also some evidence for the dispersive feature between 1.7 and 2.3eV due to a d-d* transition of $Co^{2+}$ [44].

The MCD spectrum for sample B in figure 5(b) is dominated by the strong signal below the band edge that is characteristic of a magnetic oxide, its large size occurs because of the high magnetization in this sample. The weak negative signal at $2.1 <E <3$ eV corresponds to part of the d-d* dispersive transition that has been broadened by disorder.

The spectrum of sample C, shown in figure 5 (c) was taken at remanence, it was taken after the field was reduced to zero after first saturating the sample in each direction; it is weaker than that taken in field by a factor of $M_r/M_s$ but also the d-d* transition of $Co^{2+}$ is suppressed because it is largely paramagnetic. It has a combination of the features of samples A and B. The M–G fitting works well at low energies but is unable to account for the signal near the band edge. Thus this spectrum contains the signature of both the oxide and the metallic magnetism. We have fitted the spectra to the result from equation (3) over the energy range 1.5 − 2.5 eV, where the magneto-optical response from the ZnO is expected to be negligible [42,43] using $f$, $L_{xx}$ and $\tau$ as free parameters; the fitting parameters are given in Table II. We find $f_{MCD} =(1.4\pm1)\times10^{-3}$, which may be compared with $f_{struct} \leq 2.5\times10^{-3}$ obtained from the measurements of the Co concentration and the fraction of Co ions that have a metallic environment. Again the MCD measurement, detecting the volume fraction that behaves optically like metallic Co, gives a value that is somewhat lower than that from the fraction of Co with the metallic valence of $Co^0$.

**Table II. Fitting parameters used in figures 5(a) and (c)**

|  | $L_{xx}$ | $\tau$ (eV$^{-1}$) | $f_{MCD}$ | $f_{struct.}$ |
|---|---|---|---|---|
| A: ZnCoO | 0.36 | 0.16 | $3.1\times10^{-3}$ | $7\times10^{-3}$ |
| C: ZnCoAlO | 0.4 | 0.22 | $1.4\times10^{-3}$ | $\leq 2.5\times10^{-3}$ |

We have measured hysteresis loops with MCD for sample C at the two energies, 1.9eV and 3.3eV, marked by arrows in Figure 5(c) and scaled the results so that the saturation values coincide. These energies were chosen because the contribution from the polarised Co nanoparticles dominates at 1.9eV while the ZnO matrix contribution dominates at 3.3 eV. The results are shown in figure 6. It can be seen that the coercive field is the same at both energies. This indicates that the magnetisation of the two components are coupled.

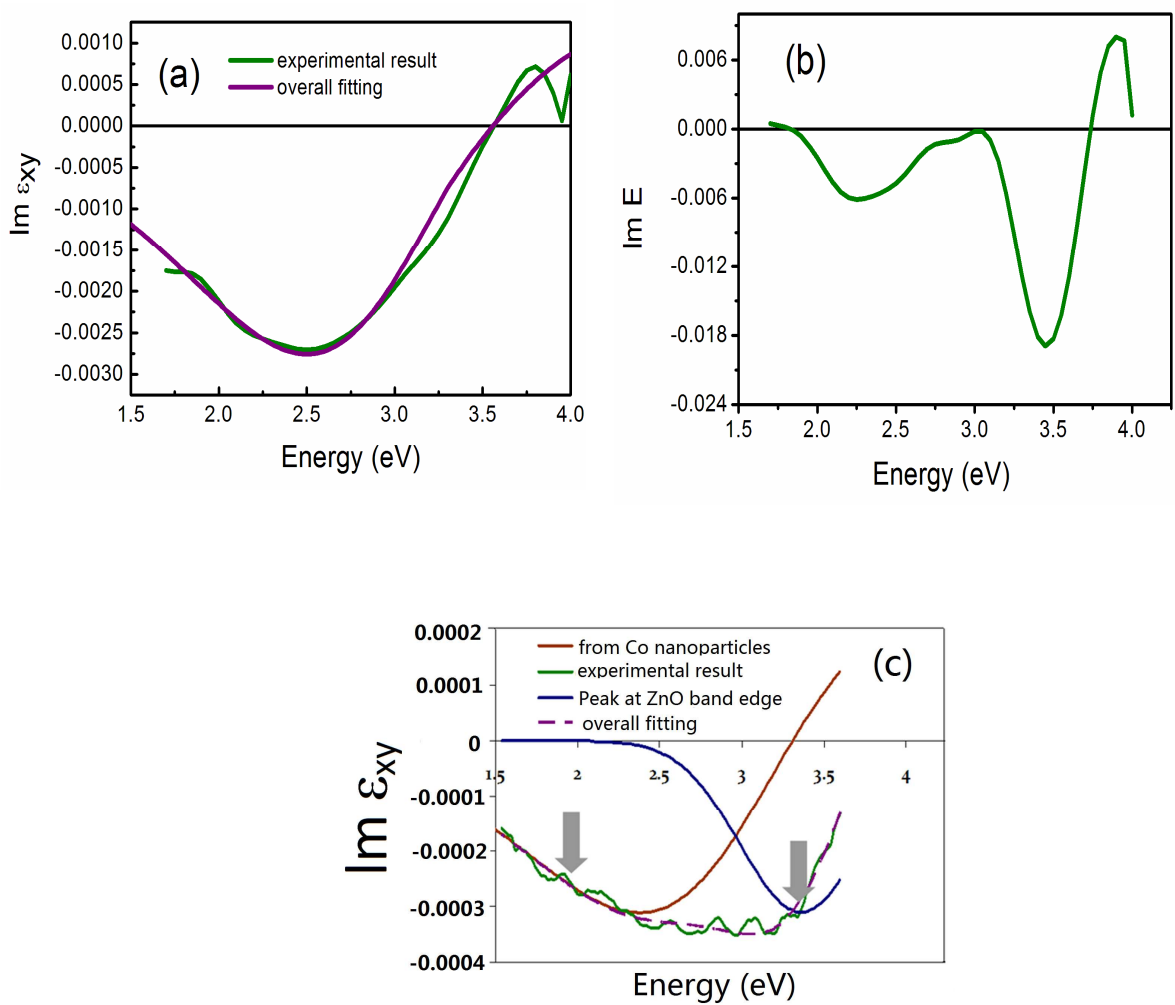

**Figure 5.** Data for $\text{Im}\,\varepsilon_{xy}(\omega)$ samples A, B and C taken at 300 K in (a), (b) and (c) respectively. The green lines shows the imaginary part of the off-diagonal dielectric tensor deduced from the MCD data. For (a) and (c) the brown line is a fit to the contribution from the metallic nanoparticles using Maxwell-Garnett theory, and the blue line shown in (c) is a Gaussian centred at 3.4 eV, which corresponds to the ZnO band edge signal. The dashed line is the sum of the two fitting curves. The data for samples A and B was taken in an applied field (12kOe) and that for sample C at remanence.

Previous measurements have been made of the hysteresis loops at temperatures between 10K and 300K obtained from the SQUID and MCD measurements made at six different energies for ZnCoAlO which was grown from the same target as sample C [45]. The coercive field shows a dramatic drop from ~1,000 Oe to ~100 Oe over this temperature range and the MCD and SQUID results remain compatible over the whole temperature range. This shows that the magnetic behaviour of the oxide is following that of the Co nanoparticles. Since the spectra from Co nanoparticles and oxide magnetism are so different, these plots indicate that the coercive field is essentially the same for both components even as it varies with temperature. This can only occur if they are exchange-coupled to each other.

## 5 Discussion

### 5.1 Estimation of Co nanoparticle size and their average separation for sample C

The size of magnetic nanoparticles, $d$, can normally be determined from the superparamagnetic blocking temperature $T_B$ via the relationship $V \approx 25 k_B T_B / K_{eff}$, where V is the average particle volume, and $K_{eff}$ is the anisotropy parameter for fcc cobalt, $K_{eff} = 5 \times 10^5$ Jm$^{-3}$ [29]. In our case the determination of $T_B$ is not straightforward, since we are dealing with a coupled system. We fitted the theory to the plot of $H_c$ as shown in figure 4(b) to obtain $T_B \sim 250$ K, which gives $V \sim 175$ nm$^3$, $d\sim$7nm. The exact value of $K_{eff}$ is somewhat uncertain since it depends on the particle size as surface, shape and strain anisotropies play a role [30].

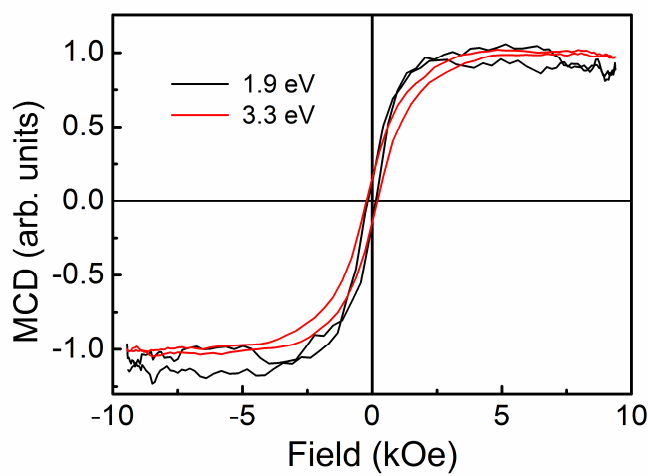

**Figure 6.** MCD hysteresis loops measured at room temperature for sample C at energies 1.9 eV and 3.3 eV. The paramagnetic contribution due to the MCD from the substrate has been subtracted from the plots and the plots scaled to the same value at saturation.

The nanoparticle separation, $R$, can be deduced from the value of $V$ and the film composition according to $R^3 = V/f$, where $f$ is the fraction of the sample occupied by the Co inclusions. Using $V \sim 175$ nm$^3$ estimated above and the value of $f = 0.0025$ deduced by PIXE and EXAFS, we arrive at $R \sim 44$nm for our sample. Knowledge of the average nanoparticle spacing and magnetization allows us to estimate the dipolar interaction energy between the clusters. For two clusters separated by ~44nm, this energy is very small (less than 1K), and so will have a negligible effect on the magnetic properties.

## 5.2 Justification of exchange coupling model

The exchange-coupling mechanism that we propose will be valid if the exchange length, $l_{ex}$, is comparable or larger than $R/2$, where $R$ is the distance between the Co nanoparticles estimated in Section 5.1 above. The exchange length can be estimated from $l_{ex} = \sqrt{A/\mu_0 M_{sat}^2}$, where $M_{sat}$ is the saturation magnetization at low temperatures, and $A$ is the exchange stiffness of the oxide matrix [46]. The latter can be estimated from $A = 0.0754 \left( M_{sat}(0)/g\mu_B \right)^{1/3} k_B \theta^*$, where $\theta^*$ characterises the rate of change of the saturation magnetization with temperature of an oxide sample in the spin-wave model according to $M_{sat}(T) = M_{sat}(0)\left(1 - (T/\theta^*)^{3/2}\right)$ [47]. We want to evaluate the exchange stiffness of the oxide magnet so we use the magnetisation data at 5K and 300K for sample B to estimate $\theta^* \approx 810$K, indicating a very high exchange-stiffness and hence a long exchange length. This, when combined with the experimental value of $M_{sat}(0)$, gives $l_{ex} \sim 21$ nm, which should be compared to $R/2 \approx 22$nm deduced from the blocking temperature and the film composition. We thus find that the estimated exchange length, $l_{ex}$, and inter-particle separation, $R$, are, indeed, comparable in this film, as required for strong exchange-coupling. This naturally leads to the determination of the material properties required for the exchange coupling to be effective. The oxide magnetism must have substantial exchange stiffness and a sufficiently high concentration of Co nanoparticles.

## 6. Conclusion

We have presented a detailed study of the magnetic and optical properties of a ZnCoAlO thin film sample that contained a significant fraction of Co nanoparticles. We deduce that we have a ferromagnetic system in which the nanoparticles contribute only a small fraction of the total magnetisation but, nevertheless, cause the *whole* sample to exhibit a large coercive field at 5K and remanence through the exchange-coupling mechanism. The presence of the Al seems to be crucial in mediating this exchange because it provides extra carriers, although coupling over a distance of 40nm through pure ZnO has also been detected recently [48]. Thus at low temperatures, where the nanoparticles are blocked, the small fraction of metallic Co is able to hold up the magnetism of the entire sample. If this were not so, the very different hysteresis loops seen at low temperatures for samples A and B would result in sample C showing a low remanence and a two component hysteresis loop at low temperatures. The results imply that a larger blocking temperature and hence a usable remanence at room temperature, could be obtained by using larger nanoclusters, provided that the concentration is sufficiently high for exchange-coupling to occur. Furthermore, the MCD data imply that the conduction band in the magnetised ZnCoAlO is spin-polarised, and hence that the system should serve as a highly useful source of spin-polarised carriers for injection into other semiconductors. The startling conclusion of this study is that incorporating controlled quantities of cobalt nanoparticles can be extremely beneficial in the search for useful oxide devices.

**Acknowledgements**

We should like to thank T. Tietze of the MPI Stuttgart, Germany for several confirmatory SQUID measurements, C. Clavero and D. Gamelin for helpful correspondence, and the EPSRC for financial support. Use of the Advanced Photon Source, an Office of Science User Facility operated for the U.S. Department of Energy (DOE) Office of Science by Argonne National Laboratory, was supported by the U.S. DOE under Contract No. DE-AC02-06CH11357.

**Appendix**

In this Appendix we describe the model that is used to generate the fits in Fig. 3. The model assumes a non-interacting system of single-domain particles where the particle diameters are given by a log-normal distribution, $f(y)$. Because of the dispersion of intrinsic particle properties, in general the particles exhibit the whole range of magnetic properties from thermal equilibrium (superparamagnetic) behaviour to thermally stable (hysteretic) behaviour; the latter giving rise to the coercivity and remanence. Taking into account the dispersion of particle volumes we can write the average magnetization, <M>, in terms of the saturation magnetization, $M_{sat}$, the external field, $H$, and the temperature as [33]:

$$\frac{<M>}{M_{sat}} = \int_0^{y_p} L(y,H,T)f(y)dy - \int_{y_p}^{y_p(H)} f(y)dy + \int_{y_p(H)}^{\infty} f(y)dy \quad (A.1a)$$

$$\text{where } f(y) = \frac{1}{\sqrt{2\pi}\sigma y} e^{-\frac{(\ln y)^2}{2\sigma^2}}. \quad (A.1b)$$

Here $y_p = D_p / D_m$ with $D_p$ the critical volume for superparamagnetic behaviour and $D_m$ the median diameter; $y_p(H)$ is the critical volume in the applied field $H$ which can be shown to be given by $y_p(H) = y_p(1 - H/H_K)^{2/3}$ where $H_K = 2K/M$ is the anisotropy field. In equation (2a) the first term on the RHS is the contribution of the superparamagnetic particles where $L(y,T,H)$ is the Langevin function for particles of metallic Co of diameter, $y$. The second and third terms represent respectively those particles which reverse in the field $H$ and those which remain stable. All the magnetic properties are derivable from equation (A.1a). The remanence and the coercive field are given respectively by,

$$M_r = \frac{M_s}{2}\left[1 - \int_{-\infty}^{x} f(y)dy\right], \quad x = \frac{1}{3\sigma}\ln\frac{T}{T_K} \quad (A.2a)$$

$$0 = \int_0^{y_p} L(y,H_c,T)f(y)dy - \int_{y_p}^{y_p(H_c)} f(y)dy + \int_{y_p(H_c)}^{\infty} f(y)dy \quad (A.2b)$$

The concentration of the nanoparticles is sufficiently low that dipolar interactions between the nanoparticles are too weak to play a role near the observed blocking temperature and so the theoretical calculations were performed assuming a log-normal distribution of non-interacting Co nanoparticles [33] with anisotropy and magnetisation appropriate for fcc cobalt, the data for $H_c$ was fitted by choosing $\sigma = 0.35$ for sample A and $\sigma = 0.1$ for sample C.

**References**

1. Schmidt G, Ferrand D, Molenkamp LW, Filip AT, van Wees BJ 2000 Phys Rev B **62** R4790

2. Pearton SJ, Abernathy CR, Norton DP, Hebard AF, Park YD, Boatner LA, Budai JD 2003 *Mat. Sci. and Eng. R* **40** 137

3. Wolf SA, Awschalom DD, Buhrman RA, Daughton JM, von Molnar S, Roukes ML, Chtchelkanova AY, Treger DM 2001 *Science* **294** 1488

4. Edmonds KW, Boguslawski P, Wang KY, Campion RP, Novikov SN, Farley NRS, Gallagher BL, Foxon CT, Sawicki M, Dietl T, Nardelli MB, Bernholc J 2004 *Phys. Rev. Lett.* **92** 037201

5. Dietl , T 2000  Science  **287** : 1019 2000

6. Sato K and Katayama-Yoshida H 2000 *Jpn. J. Appl. Phys.* **39** L555

7. Coey J M D  2006 *Current Opinion in Solid State and Materials Science* **10**,83-92

8. Venkatesan M, Stamenov P, Dorneles L S, Gunning R D, Bernoux B and Coey J M D 2007 *Appl. Phys. Lett.* **90** 242508

9. Opel M, Nielsen K-W, Bauer S, Goennenwein S T B, Cezar J C, Schmeisser D, Simon J, Mader W and Gross R 2008 *Eur. Phys. J.* B **63** 437

10. Ney A, Opel M, Kaspar TC, Ney V, Ye S,  Ollefs K, Kammermeier T,  Bauer S, Nielsen K-W, Goennenwein S T B,Engelhard M H,  Zhou S,  Potzger K, Simon J, Mader W, Heald SM, Cezar JC, Wilhelm F, Rogalev A, Gross R and Chambers SA 2010 New J. Phys. 12 013020

11. Jedrecy N, von Bardeleben H J and Demaille D 2009 *Phys. Rev.* B **80** 205204

12. Hong NH, Sakai J, Poirot N, Brize V 2006 *Phys. Rev* **73**, 132404

13. Behan A J, Mokhtari A, Blythe H J, Score D, Xu X-H, Neal J R, Fox A M and Gehring G A 2008 *Phys. Rev. Lett.* **100** 047206

14. Coey J M D, Venkatesan M and Fitzgerald C B 2005 *Nat. Mater.* **4** 173

15. Ochsenbein ST, Feng Y, Whitaker KM, Badaeva E , Liu WK, Li XS, Gamelin DR 2009 *Nat. Nanotech.* **4** 681

16. Tietze T, Gacic M, Schütz G, Jakob G, Brück S and Goering E 2008 *New J. Phys.* **10** 055009

17. Coey JMD , Stamenov P, Gunning RD, Venkatesan M , Paul K  2010 *New J. Phys.* **12** 053025

18. Yates KA, Behan AJ, Neal JR, Score DS, Blythe HJ, Gehring GA, Heald SM, Branford WR, Cohen LF 2009  Phys. Rev. B **80**   245207

19. Ney A, Ollefs K, Ye S, Kammermeier T, Ney V, Kaspar TC, Chambers SA, Wilhelm F, Rogalev A 2008  *Phys. Rev. Lett.* **100**  157201

20. Ney A, Ney V, Ye S,, Ollefs K, Kammermeier T, Kaspar TC, Chambers SA, Wilhelm F, Rogalev A 2010  *Phys Rev* B **82**   041202

21. Straumal, BB, Mazilkin, A A, Protasova SG, Myatiev AA , Straumal PB, Schutz G, van Aken PA, Goering E, Baretzky B  2009  *Phys Rev* **79**   205206

22. Kaspar T.C., Droubay T., Heald S.M., Nachimuthu P., Wang C.M., Shutthanandan V., Johnson C.A., Gamelin D.R., Chambers S.A. 2008 *New J. Phys.* **10**, 055010

23. Heald SM, Kaspar T, Droubay T, Shutthanandan V, Chambers S, Mokhtari A, Behan AJ,  Blythe HJ, Neal JR, Fox AM, Gehring GA  2009 *Phys. Rev. B*  **79**  075202

24. Schwartz DA, Norberg NS, Nguyen QP, Parker JM, Gamelin DR  2003  *J. Am.Chem. Soc.* **125**   13205

25.   Xu XH, Blythe HJ, Ziese M, Behan AJ, Neal JR,  Mokhtari A, Ibrahim RM , Fox A M and Gehring GA 2006 *New Journal of Physics* 8 135

26.  Quan ZY, Zhang X, Liu W, Li XL, Addison K, Gehring GA, and  Xu XH  2013  ACS   Appl. Mater. Interfaces **5**, 3607

27. Khalid M, Setzer A, Ziese M, Esquinazi P, Spemann D, Poppl A, Goering E 2010 *Phys Rev B* **81**   214414

28. Ying M, Blythe HJ, Dizayee W, Heald SM, Gerriu FM, Fox AM, and Gehring GA 2016 *submitted*

29.  Stoner E C and Wohlfarth  E P   1948 *Phil. Trans. R. Soc. Lond.*  A**240**  599

30.  Zhou SQ, Potzger K, von Borany J, Grotzschel R, Skorupa W, Helm M, Fassbender J 2008  *Phys.Rev.* B **77** 035209

31.Chantrell  R W,  Walmsley N, Gore J and Maylin M   2000  *Phys. Rev. B* **63**  024410

32. Gehring GA, Blythe HJ, Feng Q, Score DS, Mokhtari A, Alshammari M, Al Qahtani MS, Fox AM   2010  *IEEE Trans. on Magn.* **46** 1784-1786

33. Papusoi Jr C, Stancu Al, Dormann JL (1997)  *J Mag. Magn. Mater.* **174**  236

34. O'Grady K,  Chantrell RW,  Popplewell J and Charles S.W   1980 *IEEE Trans. Magnetics* **16**, 1077

35. Fukuma Y, Asada H, Yamamoto J, Odawara F, and Koyanagi T 2008  *Appl. Phys. Lett.* **93**, 142510

36. Clavero C, Cebollada A, Armelles G, Huttel Y, Arbiol J, Peiro F, Cornet A  2005 *Phys. Rev. B* **72**   024441

37. Clavero C, Sepulveda B, Armelles G, Konstantinovic Z, del Muro MG, Labarta A, Batlle X,   2006  *J. Appl. Phys.* **100**  074320

38. Clavero C, Armelles G, Margueritat J, Gonzalo J, del Muro MG, Labarta A, Batlle X, 2007 *Appl. Phys. Lett.* **90**  182506

39. Sun X W and Kwok H S 1999 *J. Appl. Phys.* **86** 408

40. Krinchik G S    1964   *J. Appl . Phys.*   **35**  1089

41. Score DS, Alshammari1 M, Feng Q1, Blythe HJ, Fox AM, Gehring GA, Quan ZY, Li XL, Xu XH   2010  J. *Phys.: Conf. Ser.* **200**  062024

42. Ando K, Saito H,  Zayets V and Debnath M 2004 J. Phys. Cond. Mat. **16**  S5541

43. Neal, J. R, Behan, A. J., Ibrahim, R. M., Blythe, H. J., Ziese M, Fox AM, Gehring GA  2006   *Phys Rev. Lett.* **96**  197208

44. Kittilstved K.R., Zhao J, Liu W.K., Bryan J.D., Schwartz D.A., and Gamelin D.R.(2006) *Appl. Phys. Lett.* **89**,  062510

45. Gehring G.A., Alshammari M.S., Score D.S., Neal J.R., Mokhtari A. and Fox A.M.  J. Magn. Magn. Mater. (2012) **324**,  3422

46. Skomski R. and Coey J.M.D. *Permanent Magnetism*: Studies in Condensed Matter Physics (IoP 1999)

47. Herring  C. and Kittel  C.    1951   *Phys. Rev.*   **81** 869

48. Li X., Jia J. , Guo Y., Gehring G.A. and Xu X. (2015) SPIN 5, 1540008